\pdfoutput=1

\documentclass[12pt]{iopart}

\usepackage{iopams}
\usepackage{graphicx}
\usepackage{subcaption}
\usepackage{amssymb}
\usepackage{todonotes}
\usepackage{braket}
\usepackage[normalem]{ulem}
\usepackage{xspace}
\usepackage{amsthm}
\usepackage{hyperref}
\usepackage{cite}
\usepackage{setstack} 

\newcommand{\maxthreesat}{MAX-3SAT\xspace}
\newcommand{\threesat}{3SAT\xspace}

\newcommand{\cmaxthreesat}{C_{\phi}}
\newcommand{\hmaxthreesat}{H_{\phi}}
\newcommand{\cthreesat}{\widetilde{C_{\phi}}}
\newcommand{\hthreesat}{\widetilde{H_{\phi}}}

\newcommand{\variant}[1]{\emph{Variant~#1}}
\newcommand{\step}[1]{\emph{Step #1}}

\theoremstyle{definition}
\newtheorem*{definition}{Definition}

\begin{document}

\title[Amplitude amplification-inspired QAOA]{Amplitude amplification-inspired QAOA: Improving the success probability for solving \threesat}

\author{Alexander Mandl, Johanna Barzen, Marvin Bechtold, Frank Leymann, Karoline Wild}

\address{Institute of Architecture of Application Systems, University of Stuttgart, Universitätsstraße 38,
70569 Stuttgart, Germany}
\eads{\mailto{[firstname.lastname]@iaas.uni-stuttgart.de}}

\begin{abstract}
The Boolean satisfiability problem (SAT), in particular \threesat 
with its bounded clause size, is a well-studied problem since
a wide range of decision problems can be reduced to it. Due to its 
high complexity, examining potentials of quantum algorithms 
for solving \threesat more efficiently is an important topic.
Since \threesat can be formulated as unstructured search for 
satisfying variable assignments, the amplitude amplification algorithm 
can be applied. However, the high circuit complexity of 
amplitude amplification hinders its use on near-term quantum systems.
On the other hand, 
the Quantum Approximate Optimization Algorithm
(QAOA) is a promising candidate for solving \threesat
for 
Noisy Intermediate-Scale Quantum devices in the near future
due to its simple quantum ansatz.
However, although QAOA generally exhibits a high approximation ratio, there are \threesat problem instances where its success probability decreases using current implementations.
To address this problem, in this paper 
we introduce amplitude amplification-inspired variants of 
QAOA to improve the success probability for \threesat.
For this, (i) three amplitude amplification-inspired QAOA variants are 
introduced and implemented, (ii) the variants are experimental 
compared with a standard QAOA implementation, and (iii) 
the impact on the success probability and ansatz complexity is analyzed. 
The experiment results show that an improvement in the 
success probability can be achieved with only a moderate increase 
in circuit complexity.

\end{abstract}

\maketitle

\section{Introduction}\label{sec:intro}

The Boolean satisfiability problem (SAT) is a well-studied decision problem in computer science~\cite{Kautz92, Gent94thesat, Kilby2005}.
A range of application scenarios 
from formal verification of hardware~\cite{Prasad_2005} to planning problems~\cite{Kautz92, Zhang2012} can be 
encoded as SAT instances. 
Due to its strict form and its limitation of clause size to three literals, 
often \threesat is considered, because all SAT instances can be reduced to \threesat~\cite{Arora2009}.
Various solvers for \threesat exist on classical computers.
However, since \threesat is NP-complete~\cite{Cook1971}, 
these solvers have high runtime complexity in the worst case, 
and therefore several works are currently investigating 
the possibility of quantum algorithms for solving \threesat~\cite{Akshay2020reachability, Zhang2021}.

Quantum computing promises to speed up various applications.
To implement \threesat solvers,
different quantum algorithms, such as amplitude amplification~\cite{Brassard_2002, Grover1996}
or variational approaches such as the Quantum Approximate Optimization 
Algorithm (QAOA)~\cite{Fahri2014}, have been employed.
However, current Noisy Intermediate-Scale Quantum (NISQ)
devices are characterized by a limited number of qubits and a 
high rate of gate errors~\cite{Leymann_2020}. This 
restricts the circuit size of quantum algorithms executable on current quantum hardware. 
Although amplitude amplification is optimal in its 
complexity for general 
unstructured search problems~\cite{Boyer_1998}, the 
quantum circuit that is required to solve \threesat for larger instances
exceeds the limits 
imposed by the current NISQ devices.

As variational quantum algorithms such as QAOA generally require shallower circuits,
they have the potential to be used for \threesat on NISQ devices. 
Although QAOA was originally proposed as an algorithm for the MAX-CUT problem~\cite{Fahri2014}, it uses a cost function that can be adapted to the specific problem at hand.
Thus, QAOA 
is applied to a broad range of optimization problems~\cite{Fahri2014, Wurtz_2021, Bartschi_2020, Golden_2021}.
However, while in certain applications it was experimentally 
shown that QAOA obtains a set of solutions of robust approximation ratio~\cite{Zhang2021},
the optimization procedure tends to amplify suboptimal solutions~\cite{Bennett2021}.
Thus, the original implementation of QAOA might find
wrong solutions for decision problems such as the \threesat problem.

For general optimization problems, it has been experimentally demonstrated that 
the approximation ratio of QAOA can be improved by various adjustments to 
the algorithm.
These include reformulations of the cost function~\cite{Golden_2021, Jiang_2017}, 
the used mixing operator~\cite{Bartschi_2020}, or the 
parameter initialization strategy~\cite{Egger_2021, Truger_2022, Lee_2021}.
Some of these adaptions~\cite{Bartschi_2020, Chiang2022} further make use of concepts introduced in
amplitude amplification algorithms such as Grover's search algorithm~\cite{Baritompa2005, Brassard_2002, Durr1996, Grover1996}
and variational adaptions thereof~\cite{Akshay2020reachability, Morales_2018}.
Since \threesat can also be formulated as unstructured search for satisfying assignments,
adaptions that make use of concepts from amplitude amplification might 
improve the performance of QAOA for \threesat as well.
However, so far it has not been investigated whether 
an adapted QAOA algorithm can overcome the previously described problems of QAOA for solving \threesat. 

In this paper, we experimentally investigate if adapting QAOA using concepts from 
amplitude amplification improves the 
success probability for \threesat (\Sref{sec:res_suc_prob}) and evaluate how such adaptions influence 
the computational complexity of the ansatz (\Sref{sec:res_ans_comp}). 
To this end, we implement three different variants of QAOA and compare them using randomly generated formulas.

The remainder of this paper is organized as follows:
\Sref{sec:bg} gives the required background on the 
\threesat problem and presents QAOA and amplitude amplification. \Sref{sec:motiv} covers how 
QAOA can be applied to solve \threesat and highlights 
shortcomings of the algorithm which serve as the motivation for the paper.
\Sref{sec:adapt} presents the three different adapted QAOA variants 
that are evaluated by the experiments 
described in \Sref{sec:exp}.
\Sref{sec:results} highlights the results of the 
experiments and \Sref{sec:relwork} summarizes related approaches 
and optimizations of QAOA. 
The results are discussed regarding their applicability 
in building \threesat solvers on quantum computers in 
\Sref{sec:disc}, before summarizing this work and 
presenting future research opportunities in 
\Sref{sec:conc}.

\section{Background}\label{sec:bg}

In this section, first the SAT and \threesat problem are introduced. 
It is followed by a brief description of 
QAOA. Since the adaptions of 
QAOA presented in \Sref{sec:adapt} make use of concepts of 
amplitude amplification, this algorithm is introduced at the end of this section.

\subsection{Boolean satisfiability}\label{sec:threesat}

As stated above, determining the satisfiability of Boolean formulas (SAT) is a well-studied problem
since a wide range 
of decision problems can be reduced to instances of SAT. 
Especially the \threesat variant of the problem is studied in the context of quantum algorithms due to its simple structure and convenient mapping to quantum circuits. 

The \threesat problem concerns propositional formulas $\phi = \bigwedge_{j=1}^m c_j$ in conjunctive normal form. These formulas are composed of $m$ clauses, each containing the disjunction of three literals: $c_j = \bigvee_{i=1}^3 l_{i,j}$. Each literal $l_{i,j}$ is either an atomic variable $a_k$ from the set of Boolean variables $\{a_1, \dots a_n\}$ or its negation $\lnot a_k$. The literals are referred to as \emph{negative} if its atomic variable is negated and as \emph{positive} otherwise.
To define solutions for \threesat formulas, the following definition is required:

\begin{samepage}
    \begin{definition}[\threesat interpretation]
        An \emph{interpretation} $v$ is a function that assigns $1$ or $0$ to \threesat formulas according to the following rules:
        \begin{itemize}
            \item For an atom $a_k$: $v(a_k)=0$ or $v(a_k)=1$.
            \item For a literal $l_{i,j}$ containing the atom $a_k$: 
            $v(l_{i,j}) = v(a_k)$ if the literal is positive and $v(l_{i,j}) = 1 - v(a_k)$ if the literal is negative.
            \item For a clause $c_j$: $v(c_j) = 1$ iff for one of its literals $v(l_{i,j}) = 1$ holds, and $v(c_j) = 0$ otherwise.
            \item For a \threesat formula $\phi$: $v(\phi) = \prod_{j=1}^m v(c_j)$.
        \end{itemize}
    \end{definition}
\end{samepage}

An interpretation represents the evaluation of a formula to the truth values \emph{true} and \emph{false}. Since the assignment of 
truth values to atomic variables can be represented as a binary vector, we generally refer to an 
interpretation $v$ by its variable assignment
$\vec{x} = (x_1, \dots, x_n) = (v(a_1), \dots, v(a_n))$. The evaluation of the formula according to this variable assignment is then denoted as 
\begin{equation}
\phi(\vec{x}) = \prod_{j=1}^m c_j(\vec{x}) = 1. \label{eq:threesat}
\end{equation}
Using this evaluation, the \threesat problem is given as:
\begin{samepage}
\begin{definition}[\threesat problem]
    The input is a \threesat formula $\phi$. The goal is to decide whether there exists a variable assignment $\vec{x}$ such that $\phi(\vec{x}) = 1$. Such a satisfying interpretation is referred to as a \emph{model}.
\end{definition}
\end{samepage}
\noindent Although only the question of the satisfiability of given formulas is of interest for this decision problem, a satisfying variable assignment $\vec{x}$ can usually be obtained from \threesat solvers for satisfiable instances.

\begin{figure}[t]
    \centering
    \includegraphics[width=0.7\textwidth]{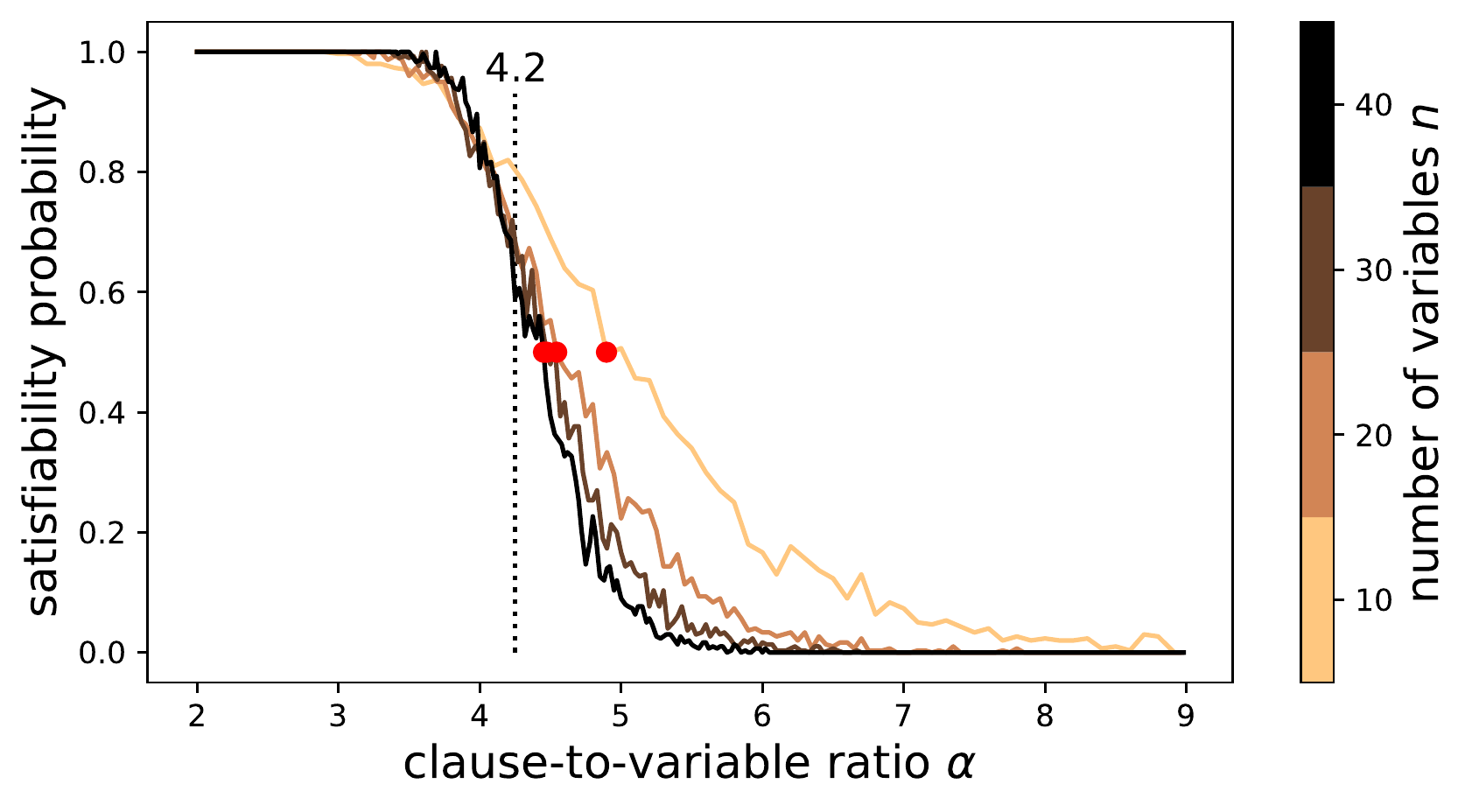}
    \caption{The proportion of satisfiable formulas for randomly generated \threesat instances, which 
    serves as an indication of the phase transition in the satisfiability probability.
    The points at each line mark the interpolated point where this proportion reaches $50\%$. As the
    number of variables $n$ increases, the points approach a clause-to-variable ratio of
    $\alpha' \approx 4.2$ as indicated by the dotted line.}
    \label{fig:phase_transition}
\end{figure}

The \threesat problem is one of the prime examples of NP-complete problems. 
Many studies were conducted to classify the 
computational resources required to solve instances of the \threesat problem 
depending on specific properties of the input formulas~\cite{Cheeseman1991, Gent94thesat, Kilby2005}. 
One of these properties is the \emph{satisfiability probability}. This is the 
probability that there exists a satisfying variable assignment for a random
\threesat formula. This satisfiability probability is 
closely linked to the \emph{clause-to-variable-ratio}
of randomly generated \threesat formulas~\cite{Cheeseman1991, Leyton_Brown_2014}.
\Fref{fig:phase_transition} shows the satisfiability probability for randomly generated formulas 
with a fixed number of variables $n$ and a varying clause-to-variable ratio $\alpha = m/n$.
The figure shows that the satisfiability probability decreases sharply 
around a clause-to-variable ratio $\alpha' \approx 4.2$~\cite{Leyton_Brown_2014, mitchell1992hard, Crawford1996}.
In this so-called \emph{phase transition} region, 
problem instances change from formulas with a small number of clauses, which are 
satisfiable with high probability, to very constrained formulas that are 
almost certainly unsatisfiable. This change in 
satisfiability probability affects the runtime complexity 
of SAT solvers. For formulas near the phase transition, the time required to solve them generally increases.
This phenomenon was experimentally shown by examining the computing time 
required for backtracking search-based solvers~\cite{Cheeseman1991}, the Davis-Putnam 
procedure~\cite{mitchell1992hard}, and similar algorithms~\cite{Crawford1996}.
Moreover, it has also been observed 
that the phase transition affects quantum algorithms for \threesat.
For formulas in the phase transition region, the probability 
that satisfying variable assignments are obtained decreases~\cite{Zhang2021, Gabor2019}.
Since the hardest instances for the \threesat problem are concentrated near
the phase transition region, this classification 
supports finding 
appropriate problem instances for testing and verifying newly devised \threesat 
solving algorithms. As such, the clause-to-variable ratio is also used to 
classify the inputs for the experiments shown in \Sref{sec:exp}.

A common characteristic to measure the quality of quantum algorithms solving \threesat
is the \emph{success probability} for randomly generated formulas.
It describes the probability that the decision (satisfiable or unsatisfiable)
that is obtained by an algorithm for a formula of a specific clause-to-variable ratio
is correct. This value will also be used in this work to inspect the 
performance of the proposed variants.

Another variant of the SAT problem is the \maxthreesat problem, which 
formulates \threesat as an optimization problem. This 
problem seeks to not specifically find 
a variable assignment $\vec{x}$ that satisfies the overall formula $\phi$, but it 
aims to give assignments that maximize the number of satisfied \threesat clauses.
These assignments can serve, for example, as heuristics for planning problems~\cite{Zhang2012}.

Using the notation from Equation \ref{eq:threesat}, the goal of \maxthreesat is to minimize
the cost function 
\begin{equation}
    \cmaxthreesat(\vec{x}) = \sum_{j=1}^m (1-c_j(\vec{x})), \label{eq:maxthreesat_cf}
\end{equation}
which counts the number of unsatisfied clauses in a given formula 
containing $m$ clauses. In contrast to the success probability that is used to evaluate algorithms 
solving \threesat, for optimization
algorithms such as \maxthreesat, the \emph{approximation ratio} is evaluated.
This is the ratio between the number of clauses that are satisfied by a variable assignment that is
obtained from an optimization algorithm and the number of clauses that are satisfied 
in an optimal assignment. For a given \threesat formula $\phi$, any optimal assignment 
for the \maxthreesat problem satisfies $m_{opt} \leq m$ clauses. If $\phi$ is not 
satisfiable, $m_{opt} < m$ holds. For an arbitrary variable assignment $\vec{x}$
that satisfies $m_{\vec{x}}$ clauses, the approximation ratio is given as 
$m_{\vec{x}}/m_{opt}$. 
Currently, a lot of research in the field of 
quantum algorithms focuses on improving the approximation ratio for optimization problems 
such as \maxthreesat using a quantum computer. One of the algorithms that are applied to achieve this 
goal is QAOA.

\subsection{The Quantum Approximate Optimization Algorithm}\label{sec:qaoa}

The Quantum Approximate Optimization Algorithm (QAOA) is a variational quantum 
algorithm designed to solve combinatorial optimization problems and was 
first proposed as a quantum solution for the MAX-CUT problem~\cite{Fahri2014}. 
This algorithm uses an ansatz that is modeled 
for the specific problem at hand. This allows it to also be extended to 
other problems, including some commonly known NP-complete problems~\cite{Lucas_2014}.

At its core, QAOA aims to find the states with minimal eigenvalue of a given Hamiltonian $H_C$, 
which encodes the cost function $C : \{0,1\}^n \rightarrow \mathbb{R}$ associated 
with a combinatorial optimization problem. 
For a solution candidate given by the binary vector $\vec{x} \in \{0,1\}^n$, 
$H_C$ encodes the cost of this solution in its eigenvalues as
\begin{equation}
    H_C\ket{\vec{x}} = C(\vec{x})\ket{\vec{x}}.
\end{equation}
Therefore, the eigenvalues are minimal for optimal solutions with respect to $C(\vec{x})$.
The ansatz consists of a circuit preparing an initial state, followed
by $p$ repetitions that alternately apply the \emph{phase-separation} operator
$U(H_C,\gamma_i) = e^{-i\gamma_i H_C}$ and the \emph{mixer} circuit $U(H_B, \beta_i) = e^{-i \beta_i H_B}$.
The phase-separation circuit introduces a not-measurable phase that encodes the cost of the solutions. The mixer circuit is used to translate this phase into an adjustment of the measurement probability of the solution candidates.
The unitary operations $U(H_C,\gamma_i)$ and $U(H_B,\beta_i)$ are parametrized by the real-valued vectors $\vec{\gamma} = (\gamma_1, \dots, \gamma_p)$
and $\vec{\beta} = (\beta_1, \dots, \beta_p)$.
Applying this procedure to the initial state $\ket{+}^{\otimes n}$ takes the 
quantum system to the parametrized state 
\begin{equation}
    \ket{\vec{\gamma}, \vec{\beta}} = \left( \prod_{i=1}^p U(H_B, \beta_i) U(H_C, \gamma_i) \right) \ket{+}^{\otimes n},
\end{equation}
with the mixer Hamiltonian given as $H_B = \sum_{i=1}^n X_i$, which is often referred to as the 
\emph{transverse-field} mixer~\cite{Jiang_2017}.

By executing the described quantum circuit, a set of binary vectors representing solution candidates
to the optimization problem is sampled and their average cost is calculated.
In the variational loop, a classical optimizer is used 
to find optimal parameters $\vec{\gamma}^*, \vec{\beta}^*$ such that the expected cost of the measured 
solutions is minimized. In other words, the algorithm computes 
\begin{equation}
    \min_{\vec{\gamma}, \vec{\beta}} \braket{\vec{\gamma},\vec{\beta}|H_C|\vec{\gamma},\vec{\beta}}.
\end{equation}
Using the optimal parameters $\vec{\gamma}^*, \vec{\beta}^*$ that are found by the classical optimizer,
a set of solutions that approximately minimizes the problem cost function $C(\vec{x})$ can be 
measured.  
Assuming that the number of repetitions $p$ of $U(H_B, \beta_i)$ and $U(H_C, \gamma_i)$,
is sufficiently large,
such a set of solutions should be found~\cite{Fahri2014}.
However, giving estimates for the required $p$ is still 
an actively studied question.
This value is also referred to as the $\emph{QAOA depth}$ of the ansatz as it 
directly influences the circuit depth of the executed quantum circuit.

\subsection{Amplitude amplification}\label{sec:grover}
Grover's algorithm~\cite{Grover1996}, the first algorithm making use 
of the concept of amplitude amplification, performs unstructured search 
for a unique solution state identified by an oracle on a quantum computer. 
It speeds up this process relative to classical unstructured search by requiring 
$O(\sqrt{N})$ evaluations of the oracle as opposed to $O(N)$ on a classical 
computer for a search space containing $N$ elements~\cite{Grover1996}.
It further was shown that this algorithm is optimal in its runtime complexity up to a constant factor 
for unstructured search problems~\cite{Boyer_1998}.
The idea behind this process is generalized to problems with 
multiple solution states in the amplitude amplification algorithm~\cite{Brassard_2002}.

Similar to the phase-separation operator in QAOA, amplitude amplification requires an oracle that manipulates the phase of a state based on
its membership in the set of solutions. Given some Boolean indicator function $f: \{0,1\}^n \to \{0,1\}$, 
the Grover oracle, sometimes referred to as the \emph{phase-flip oracle},
maps each candidate state $\ket{\vec{x}}$ to $(-1)^{f(\vec{x})}\ket{\vec{x}}$.
To translate this change in phase into a measurable amplification of the 
amplitude of a solution state, the \emph{Grover diffusion} operator $2\ket{+}\bra{+}^{\otimes n} - I^{\otimes n}$ is used~\cite{nielsen2002quantum}.
This process is repeatedly applied to the starting state $\ket{+}^{\otimes n}$, which 
represents the superposition of all solution candidates in an unconstrained search space. 
Depending on the size of the set of solutions 
and the size of the search space, a fixed limit can be specified for the number of applications of the oracle 
and the diffusion operator before the desired 
solution states are measured with certainty~\cite{nielsen2002quantum, Brassard_2002}.

Solution candidates for problems in NP, such as \threesat,
can be verified in polynomial time classically
and therefore can also be verified using a polynomial-size quantum circuit~\cite{nielsen2002quantum}.
Thus, a general unstructured search algorithm such as amplitude amplification is predestined 
to solve satisfiability problems. For the \threesat problem, 
a solution candidate is a variable assignment represented by a bitstring $\vec{x} \in \{0,1\}^n$.
The purpose of the phase oracle is then to mark assignments in the 
superposition of variable assignments if and only if they satisfy a given formula.

\section{Motivation and Research Questions}\label{sec:motiv}

The drawback of amplitude amplification algorithms for current NISQ devices is its requirement for deep quantum circuits.
Due to QAOA's shallower quantum circuits, the potential of applying 
QAOA to the \threesat problem is actively studied. However, since QAOA 
is an optimization algorithm, currently only the optimization variant 
of \threesat called \maxthreesat, is considered~\cite{Akshay2020reachability, Zhang2021}.
For applying QAOA to \maxthreesat, the cost function in \Eref{eq:maxthreesat_cf} is encoded 
in a cost Hamiltonian $\hmaxthreesat$.
There are various formulations of this Hamiltonian~\cite{Lucas_2014, Nuesslein2022, Chancellor2016, Herrman2021, Whitfield2012},
with some focusing on reducing the number of operations required when implementing 
this cost function in the phase-separation operator. 
The formulations generally describe the cost Hamiltonian as a sum of clause operators
\begin{equation}
    \hmaxthreesat = \sum_{j=1}^m H_{C_j}.
\end{equation}
Each $H_{C_j}$ increases the eigenvalue of the variable assignments that do not satisfy the $j$-th clause.
These clause operators are expressed as 
\begin{equation}
    H_{C_j} = \frac{1}{8}\prod_{i=1}^3 (I + p_{i,j} Z_{a_{i,j}}). 
    \label{eq:clauseham}
\end{equation}
In this equation, $p_{i,j}$ describes the polarity of the literal $l_{i,j}$ in the $j$-th clause 
with $p_{i,j} = 1$ for positive literals and $p_{i,j} = -1$ for negative literals.
Furthermore, the expression $Z_{a_{i,j}}$ describes that the Pauli operator $Z$ is 
applied to the qubit representing the atom of literal $l_{i,j}$.
Depending on the polarity of the literal $l_{i,j}$, the individual factors in \Eref{eq:clauseham} have eigenvalue $0$ if the literal is satisfied by the variable assignment.
The factors have eigenvalue $2$ if the literal is not satisfied.
Thus, if at least one literal in the disjunction is satisfied, the clause operator $H_{C_j}$ has eigenvalue $0$ for this variable assignment.
By using QAOA to find the states with minimal eigenvalue for the cost Hamiltonian $\hmaxthreesat$, the algorithm will therefore find variable assignments that have eigenvalue $0$ for as many clause operators $H_{C_j}$ as possible. 
Thus, this variable assignment maximizes the number of satisfied clauses, which solves the \maxthreesat problem. For details on how this cost operator is implemented as a quantum circuit see \ref{sec:app1}.

To extend this algorithm for \maxthreesat to solve the decision problem \threesat, 
recent works focus
on using the variable assignments that are obtained for \maxthreesat and 
inferring whether they satisfy the given input~\cite{Zhang2021, Akshay2020reachability}. 
This is achieved by evaluating the formula classically using the sampled solutions.

Using experimental evaluations, it was found that QAOA 
exhibits a good approximation ratio for \maxthreesat~\cite{Zhang2021}.
However, by reasoning from the approximated solution of the \maxthreesat problem, 
incorrect solutions are often obtained for the \threesat decision problem.
Considering a satisfiable \threesat formula in conjunctive normal form, a measured variable 
assignment satisfying 99\% of the given clauses would be a respectable 
result for an algorithm approximating the MAX-3SAT problem.
However, this would be evaluated to the wrong
result of \emph{unsatifiable} for the decision problem variant.
Similar observations were made by Bennett and Wang \cite{Bennett2021}. In their work, they 
found that the Quantum Walk Optimization Algorithm (QWOA), a generalization of 
QAOA, tends to amplify suboptimal solutions to increase the approximation ratio
as they are more numerous than optimal solutions. 
Thus, the standard QAOA implementation leaves room for improvement when 
applied to \threesat.

Therefore, the objective of this work is to exploit the advantages of 
amplitude amplification to improve the success probability of the QAOA for \threesat.
Thus, we formulate the following research questions:
\begin{samepage}
\begin{itemize}
    \item RQ1: Can mechanisms found in amplitude amplification be used to 
improve the success probability of QAOA for the \threesat problem?
\end{itemize}
\begin{itemize}
    \item RQ2: How does adapting the QAOA ansatz using concepts from amplitude 
    amplification influence the 
computational complexity of the ansatz?
\end{itemize}
\end{samepage}

\section{Amplitude amplification-inspired QAOA variants}\label{sec:adapt}

\begin{figure}[t]
    \includegraphics[width=\textwidth]{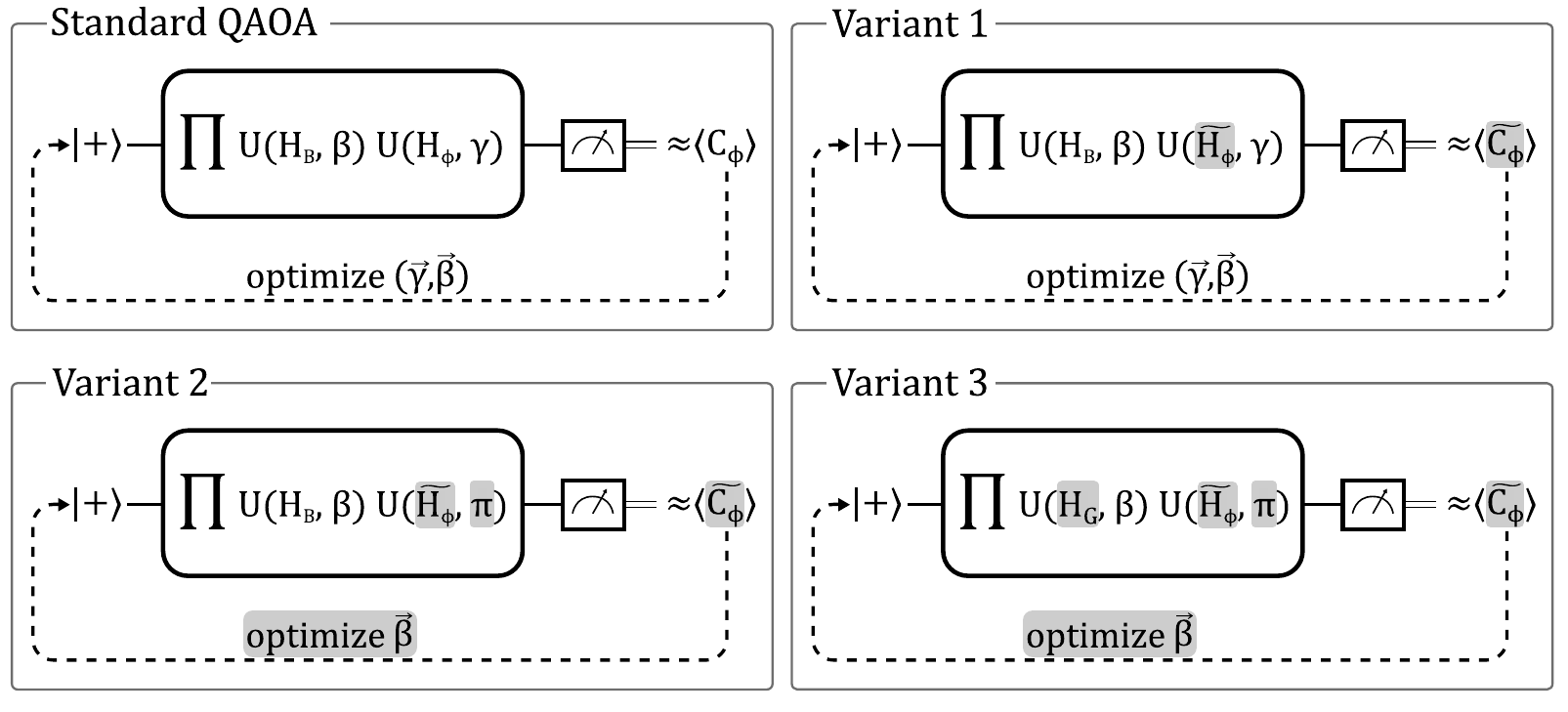}
    \caption{A schematic overview of the three different variants of QAOA implemented in this work
    and the standard QAOA implementation for \maxthreesat.
    The adaptions applied to the standard QAOA implementation for 
    the \maxthreesat problem are highlighted.}
    \label{fig:variants}
\end{figure} 

To answer the research questions, we
examine and experimentally compare three different variants of QAOA that make use 
of ideas employed in amplitude amplification
to improve the success probability for the \threesat problem. We present the 
adaptions as a gradual reformulation of QAOA until it almost matches 
the amplitude amplification algorithm.
\Fref{fig:variants} gives an overview of standard QAOA and the proposed adaptions we implement for the experiments in \Sref{sec:exp}:

\begin{itemize}
    \item \variant{1}: An adaption of QAOA that optimizes using a binary cost function.
    \item \variant{2}: An extension of \variant{1}, where the parameters $\vec{\gamma}$ are fixed to $\gamma_i = \pi$. The classical optimization step 
    will only optimize the mixer parameters $\vec{\beta}$.
    \item \variant{3}: An extension of \variant{2}, that employs 
    the Grover mixer.
\end{itemize}
In the following, these variants of QAOA are explained in detail.

\subsection{\variant{1}: Cost function and phase-separation operator}\label{sec:cf}

\variant{1} applies an adapted binary cost function. To guide the 
approximation algorithm away from amplifying suboptimal solutions, this function
penalizes all but optimal solutions equally. Given some \threesat 
formula $\phi$, this objective function $\cthreesat(\vec{x})$ is given as 
\begin{equation}\label{eq:cf_threesat}
    \cthreesat(\vec{x}) = 
    \cases{
        0 & if $\phi(\vec{x}) = 1$\\
        1 & if $\phi(\vec{x}) = 0$\\
    },\label{eq:threesatcf}
\end{equation}
where $\phi(\vec{x}) = 1$ represents that the formula $\phi$ evaluates to 
true given the variable assignment $\vec{x}$
and $\phi(\vec{x}) = 0$ representes that the formula evaluates to false.
Similarly, the corresponding cost Hamiltonian $\hthreesat$ has eigenvalue~$1$
 for quantum states $\ket{\vec{x}}$ representing 
unsatisfying assignments and eigenvalue~$0$ for satisfying assignments. 
The expected cost $\braket{\vec{\gamma}, \vec{\beta}|\hthreesat|\vec{\gamma}, \vec{\beta}}$ 
approximated by QAOA using this operator 
can therefore be seen as the proportion of unsatisfying assignments that are measured 
at the end of the circuit.

We implement the phase-separation operator for QAOA based on this cost function 
by reusing existing techniques for oracle synthesis~\cite{Schmitt2022}. We first present the general approach that 
is used to extend 
a bit-flip oracle to a phase-separation operator as it is used for our experiments. A general 
bit-flip oracle for a given function \mbox{$f: \{0,1\}^n \to \{0,1\}$} is defined as 
\begin{equation}
    U \ket{\vec{x}, y} = \ket{\vec{x}, y \oplus f(\vec{x})}.
\end{equation}
Thus, it encodes the result $f(\vec{x})$ by negating an ancilla qubit. This qubit can 
further be used to manipulate the phase of the examined quantum state as it is 
required in QAOA.

We perform this extension by noting that the required phase-separation operator $U(H_C, \gamma)$
takes $\ket{\vec{x}}$ to $e^{-i\gamma C(\vec{x})} \ket{\vec{x}}$ for a given assignment of 
combinatorial variables $\vec{x}$ and an arbitrary objective function. 
Assuming the cost function is binary and given as $C(\vec{x}) \in \{0,1\}$,
this exact operation can 
also be performed using a bit-flip oracle, by applying
the phase gate $P(-\gamma)$ before uncomputing:
\begin{equation}\label{eq:oracle_ext}
    \eqalign{
    \ket{\vec{x}, 0 \oplus C(\vec{x})} &= \ket{\vec{x}, C(\vec{x})}\\
    &\overset{P(-\gamma)}{\longrightarrow} e^{-i\gamma C(\vec{x})}\ket{\vec{x}, C(\vec{x})}\\
    &\overset{U^\dagger}{\longrightarrow} e^{-i \gamma C(\vec{x})} \ket{\vec{x}, 0}.} \label{eq:cftophase}
\end{equation}
For the \threesat problem, the cost function $\cthreesat(\vec{x})$ in \Eref{eq:threesatcf} is binary. 
We therefore perform the process in \Eref{eq:cftophase} using an oracle circuit that flips the ancilla qubit 
if the given formula is satisfied. To obtain a phase-separation operator that 
matches $U(\widetilde{H_\phi}, \gamma)$ and only introduces the phase if the 
formula is not satisfied, the ancilla qubit is negated before the oracle is applied.

\subsection{\variant{2}: Phase-flip oracle}\label{sec:pf_oracle}

With the adaption of \variant{1}, first parts of amplitude amplification were 
incorporated into QAOA: The phase-separation operator based on a bit-flip oracle as presented above 
can be seen as a phase-flip oracle with an additional degree of freedom represented by 
the parameter $\gamma$.
By fixing $\gamma = \pi$, the phase-flip oracle is obtained 
since $e^{-i \pi C(\vec{x})}\ket{\vec{x}} = (-1)^{C(\vec{x})}\ket{\vec{x}}$. 
Additionally, fixing this parameter simplifies the optimization loop, since 
then only the optimization over the mixer parameters $\vec{\beta}$ remains. We therefore 
evaluate the variant with the parameter $\gamma$ fixed to $\gamma=\pi$ as \variant{2} 
in our experiments.

\subsection{\variant{3}: Mixer and Grover diffusion}

Similar to QAOA, amplitude amplification can be described as the repeated application of 
a gate that encodes a cost function (the phase-flip oracle) and a mixing gate (in the context of 
amplitude amplification referred to as the diffusion operator). The similarity between 
the phase-flip oracle and QAOA's phase-separation operator has already been sketched in 
\Sref{sec:pf_oracle}. In \variant{3}, we further adapt QAOA by using
the Grover mixer (GM-QAOA~\cite{Bartschi_2020}), which closely resembles the diffusion operator
in amplitude amplification.

Instead of the usual mixing Hamiltonian as presented in \Sref{sec:qaoa}, 
the general Grover mixer uses the Hamiltonian $H_F = \ket{F}\bra{F}$, where 
$F$ is a set of feasible solution states and $\ket{F}$ is the equally weighted superposition 
of these states~\cite{Bartschi_2020}. Since for \threesat, the objective function decides whether
a solution is feasible, the set $F$ is assumed to be all possible $n$ qubit binary 
strings representing assignments of $n$ Boolean variables. This implies $\ket{F} = \ket{+}^{\otimes n}$ and results in the mixing Hamiltonian
$H_G = \ket{+}\bra{+}^{\otimes n}$.

By applying the reformulation of the resulting mixing unitary as presented in~\cite{Bartschi_2020}, 
it is possible to show how this gate can be implemented as a quantum circuit.
In the following equations, we omit the register size $n$ to aid readability.
Using the infinite sum for the matrix exponential, the mixing unitary is reformulated as
\numparts \begin{eqnarray} 
    U(H_G, \beta) &= e^{-i \beta \ket{+}\bra{+}}\\
    &= \sum_{k=0}^{\infty} \frac{1}{k!}(-i\beta)^k \left(\ket{+}\bra{+}\right)^k.\label{eq:gmixerform1}
\end{eqnarray} \endnumparts
For $k=0$, the matrix on the right-hand side reduces to 
$\left(\ket{+}\bra{+}\right)^0 = I$. Furthermore since $\left(\ket{+}\bra{+}\right)^2 = \ket{+}\braket{+|+}\bra{+} = \ket{+}\bra{+}$, this matrix is idempotent. Therefore, $\left(\ket{+}\bra{+}\right)^k = \ket{+}\bra{+}$ is used for $k>1$ in \Eref{eq:gmixerform1} to give
\numparts \begin{eqnarray} 
    U(H_G, \beta) &= I + \sum_{k=1}^{\infty}\frac{1}{k!} (-i\beta)^k \ket{+}\bra{+}\\
    &= I + \left(e^{-i \beta} - 1\right)\ket{+}\bra{+} \label{eq:reform_grover_mixer_equiv} \\
    &= H\left(I + \left(e^{-i \beta} - 1\right)\ket{0}\bra{0}\right)H\\
    &= HX\left(I + \left(e^{-i \beta} - 1\right)\ket{1}\bra{1}\right)XH\\
    &= HX\; \mathit{CP}_{n}(-\beta)\;XH,\label{eq:reform_grover_mixer_layers}
\end{eqnarray} \endnumparts

\begin{figure}
    \centering
    \includegraphics{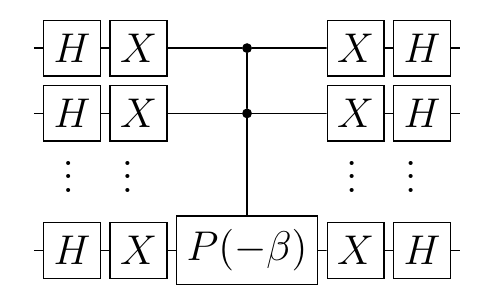}
    \caption{The quantum circuit for the Grover mixer~\cite{Bartschi_2020} with the mixing Hamiltonian $H_G = \ket{+}\bra{+}^{\otimes n}$.}
    \label{fig:grover_mixer_circuit}
\end{figure}

\noindent where $\mathit{CP}_{n}(-\beta)$ is the phase-shift gate on the $n$th qubit, which is controlled by the other $n-1$ qubits.
 Therefore, the Grover mixer can be built using two layers of Hadamard and $X$ gates and a controlled phase shift gate, resulting in the quantum circuit in \Fref{fig:grover_mixer_circuit}.
\Eref{eq:reform_grover_mixer_equiv} further highlights the relationship between 
the Grover mixer and the diffusion operator. It shows that by setting $\beta = \pi$, the Grover mixer 
is equal to the diffusion operator $2\ket{+}\bra{+} - I$ up to a global phase, which gives  
the justification for its name.

\section{Experiment setup} \label{sec:exp}

\begin{figure}[t]
    \centering
    \includegraphics[width=0.9\textwidth]{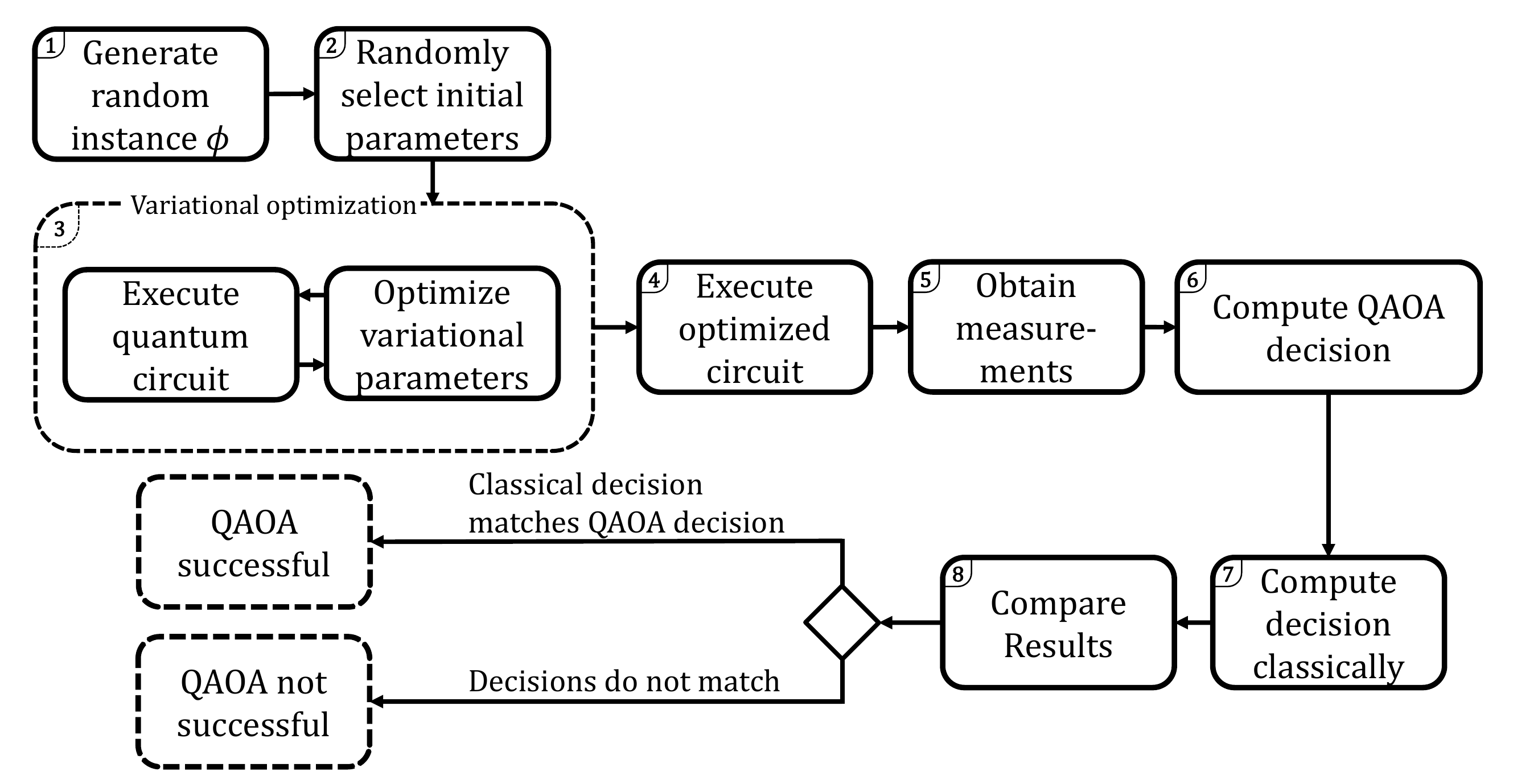}
    \caption{Experiment setup to approximate success probability. This experiment is executed 
    for \variant{1-3} as well as the baseline implementation.}
    \label{fig:exp_setup}
\end{figure}

To answer RQ1, we implement the presented variants of QAOA 
using Qiskit
and evaluate their success probability when solving \threesat. 
For this evaluation, we use Qiskit's simulator to obtain noise-free results.
This section describes both the inputs for these experiments 
and the evaluation of the results in detail.
We further implement and execute this experiment for an unmodified
\threesat solver based on QAOA for the \maxthreesat problem as described in~\Sref{sec:motiv}
as a baseline.
To further evaluate how the QAOA depth influences the success probability of each variant, 
all experiments are executed for $p \in \{2,4,6,8,10\}$. 
For each variant and each selected QAOA depth, a set of \threesat formulas is generated 
and the proportion of successful executions of the variants on these inputs is computed.
There is a tradeoff between the used instance sizes and the computation time that is required for the simulation. 
We therefore keep the number of variables at $n=10$. To be able to study the effect of the \threesat phase transition on the success probability, we evaluate formulas of clause-to-variable ratio $1.5 \leq \alpha \leq 8$. 
This way, formulas on both sides of the phase transition region at $\alpha \approx 4.2$ are included.
Thus, for 
each number of clauses $m$ between $m=15$ and $m=80$, $400$ \threesat instances are evaluated
by executing the process shown in \Fref{fig:exp_setup}.

In \step{1} of the experiment, a \threesat formula is 
randomly generated. Its clauses are composed by picking three of the 
$n=10$ available atomic variables with repetition.
The variables are added to the overall disjunction as positive or negative literals 
with equal probability. In \step{2} and \step{3}, the quantum algorithm is executed. 
The parameters are initialized and then optimized in the variational loop.
As the classical optimizer for this experiment, we choose the implementation 
of the linear approximation-based optimizer COBYLA, provided in SciPy~\cite{Virtanen2020_scipy}. 
This optimizer was found to be preferable in low-noise scenarios~\cite{Lavrijsen2020}, which is preferrable since we simulate the quantum circuits without noise.

For \variant{1-3}, the variational loop finds values for the parameters 
$\vec{\gamma}^*$ and $\vec{\beta}^*$ (or only $\vec{\beta}^*$ for \variant{2} and \variant{3} respectively) that
minimize the expected value of the cost function $\cthreesat(\vec{x})$ from 
\Eref{eq:cf_threesat}. For the baseline implementation, \step{3} minimizes $\cmaxthreesat(\vec{x})$ from \Eref{eq:maxthreesat_cf}.
The ansatz is then executed again using these optimized parameters 
 in \step{4} to measure a set of 
variable assignments in \step{5}.
From these measurements, the decision (\emph{satisfiable} or \emph{unsatisfiable}) for the \threesat problem instance is obtained 
in \step{6}. The input formula is deemed satisfiable if and only if 
\begin{equation}
    \braket{\vec{\gamma}^*, \vec{\beta}^*|\hthreesat|\vec{\gamma}^*, \vec{\beta}^*} < 1.
\end{equation}
In other words, if the measurement using the optimal parameters produces 
at least one satisfying assignment, the formula is satisfiable.
It decides the contrary if no such solution is found.
Finally, in \step{7} the exact decision for the given input is calculated classically. 
We use
the classical solver PySAT~\cite{Ignatiev2018_pysat} for this purpose.
Based on this result, one execution of the algorithm is deemed successful in \step{8} if the 
decision of the studied variant matches the decision 
obtained by the classical \threesat solver.

The presented variants of QAOA 
employ rather computationally expensive 
oracles and mixers. The number of operations in the quantum circuits is too large 
to be executed on current devices for $n=10$.  Thus, we perform the experiments 
using 
Qiskit's built-in simulation capabilities. For simplicity, we use 
the classical function compiler
in Qiskit, which makes use of tweedledum~\cite{Schmitt2022}
to obtain the required bit-flip oracles for \mbox{\variant{1-3}}.
To obtain the measurements for the 
baseline implementation we use Qiskit's built-in QAOA implementation~\cite{Qiskit_QAOA}.
Although the baseline implementation
optimizes with respect to the function $\cmaxthreesat$,
the final decision in \step{6} is still computed using $\cthreesat$ as for \mbox{\variant{1-3}}.
This allows us to compare the results 
using the same metric although different operators are optimized.

To further investigate the influence of the used optimizer on the success probability, we extend this evaluation by comparing different optimization procedures. 
We compare the optimizer that is used in this work (COBYLA) to different optimizers that are used in related works concerning variational quantum algorithms~\cite{Zhang2021, Akshay_2021, Gacon2021}: the limited memory BFGS method~\cite{Nocedal1999} and the Simultaneous Perturbation Stochastic Approximation (SPSA) algorithm~\cite{Spall1992}. 
For this comparison, we investigate the success probability of the baseline implementation and \variant{1-3} using formulas of clause-to-variable ratio $\alpha = 4.5$. 
This value is chosen since our initial experiments found that formulas of this clause-to-variable ratio experience the lowest success probability for $n=10$. 
Lastly, the initial set of parameters that are chosen for the optimization (see \step{2}) might influence the success probability of the inspected variants. 
Therefore, we also investigate the case where the optimization procedure is repeated five times with different sets of initial parameters and the result with the lowest cost is selected.

\section{Experiment results}\label{sec:results}

The results of the conducted experiments were analyzed in order 
to answer the research questions of this paper. The experiments 
have shown that the adaptions of QAOA presented in this paper 
influence the success probability positively (RQ1), but also 
increase the complexity of the quantum circuits (RQ2). The results are presented in detail below.

\begin{figure}[t]
    \centering
    \includegraphics[width=0.65\textwidth]{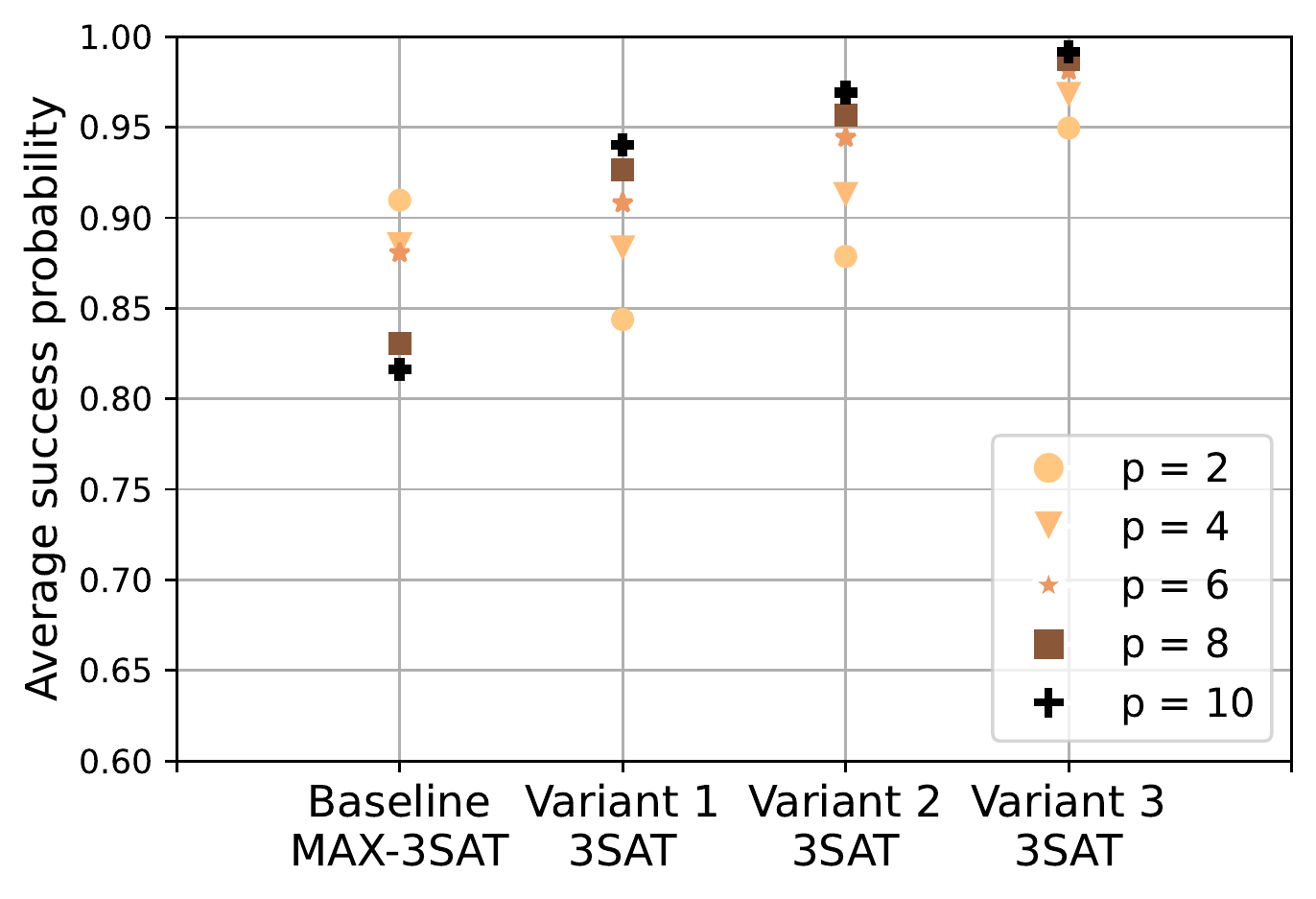}
    \caption{The success probability of the examined \threesat solver variants for the examined 
    QAOA depths $p$, over the range $[1.5, 8.0]$ of clause-to-variable ratios.}
    \label{fig:avg_probs_overall}
\end{figure}
\clearpage
\subsection{Effect on the success probability}\label{sec:res_suc_prob}

We first present the success probability of the different QAOA variants in our experiments.
The implementations used for these experiments and the obtained experimental data are available online\footnote{\url{https://github.com/UST-QuAntiL/aa_inspired_qaoa}}.
\subsubsection{Comparison of QAOA variants}

\Fref{fig:avg_probs_overall} shows the average success probability of \variant{1-3}
as well as the baseline implementation. These numbers were obtained by computing
the proportion of successful executions of the process in \Fref{fig:exp_setup} over the whole
range $[1.5, 8.0]$ of examined clause-to-variable ratios.
\Fref{fig:avg_probs_overall} shows that the rewrite of the cost function to 
encode the decision problem in the phase-separation operator 
\emph{(Variant~1)} alone only 
slightly improves the success probability for \threesat 
when higher depth ansätze of $p=8$ and $p=10$ are used. 
Our experiments even suggest that the variational optimization of the 
parameters $\vec{\gamma}$ is not necessary: \variant{2} exhibits a 
higher success probability for p=8 and p=10 than both, \variant{1} and the baseline implementation. 
Finally, the introduction of the Grover mixer in \variant{3}
has the most noticeable positive effect on the success probability as it 
improves the success probability to at least $0.94$ even for the smallest studied depth $p=2$.

\begin{figure}[t]
    \centering
    \includegraphics[width=1.0\textwidth]{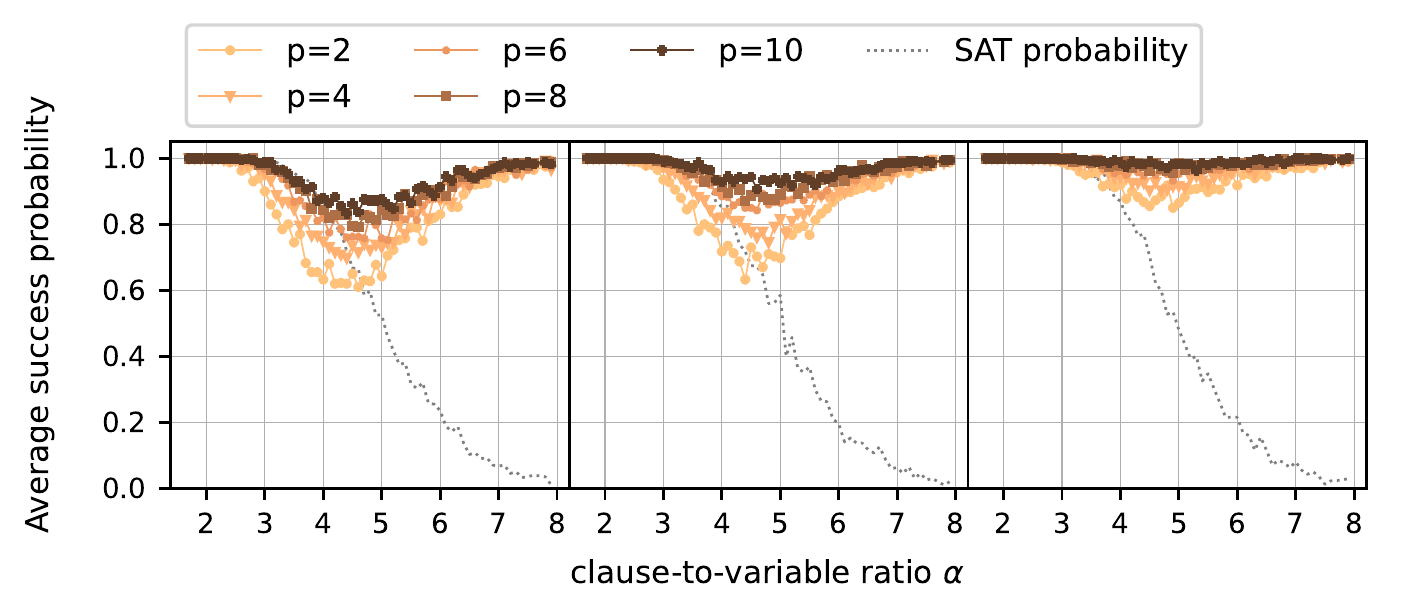}
    \hspace*{\fill}
    \begin{subfigure}[b]{0.3\textwidth}\caption{\variant{1}}\label{}\end{subfigure}
    \begin{subfigure}[b]{0.3\textwidth}\caption{\variant{2}}\label{}\end{subfigure}
    \begin{subfigure}[b]{0.3\textwidth}\caption{\variant{3}}\label{fig:exp_standard_3}\end{subfigure}
    \caption{The success probability for the examined variants for $n=10$ variables. The probability 
    is shown for differing clause-to-variable ratio $\alpha$ and various depths $p$.}
    \label{fig:exp_standard}
\end{figure}

To highlight the dependence of the results on the satisfiability probability and its 
phase transition, \Fref{fig:exp_standard} shows the obtained success probability 
of the presented variants for  
each examined clause-to-variable ratio of the inputs. 
The satisfiability probability of these inputs is computed classically and 
is shown as a dotted 
line in each subplot. For all variants, the plot shows that the hardest instances 
for QAOA concentrate near the phase transition region. The success probability 
decreases for problem instances between $\alpha=4$ and $\alpha=5$.
However, \Fref{fig:exp_standard_3} shows that \variant{3} is the least affected 
by this phenomenon. Even for the hardest instances in the phase transition 
region, the success probability only decreases to $0.8$ for the shallowest 
circuits of QAOA depth $p=2$.

\subsubsection{Comparison of optimizers}\label{sec:compo_opt}

\begin{figure}[t]
    \begin{subfigure}{1\textwidth}
        \refstepcounter{subfigure}\label{fig:optimizer_compare:cobyla}
        \refstepcounter{subfigure}\label{fig:optimizer_compare:bfgs}
        \refstepcounter{subfigure}\label{fig:optimizer_compare:spsa}
    \end{subfigure}%
    \centering
    \includegraphics{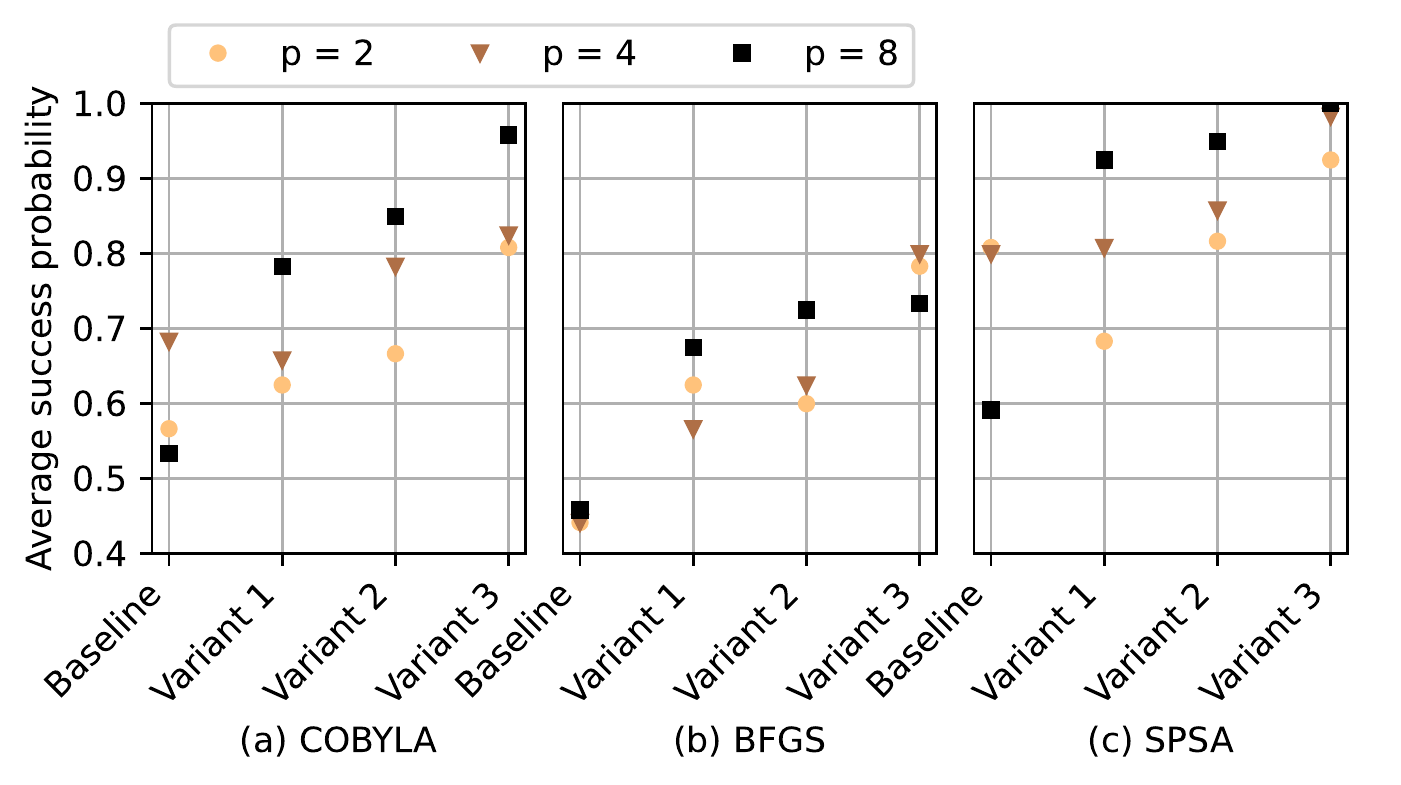}
    \caption{The success probability for the examined variants for the optimizers COBYLA, BFGS and SPSA and for different QAOA depths $p$ and \threesat formulas with $\alpha = 4.5$.}
    \label{fig:optimizer_compare}
\end{figure}

The improvement obtained by \variant{2} suggests that the optimization procedure 
is a limiting factor in the performance of the algorithm. Because \variant{2} omits 
half of the variational parameters by using the phase-flip oracle from amplitude 
amplification, it overcomes this limitation to some extent.
Using the ansatz from \variant{2}, the optimizer needs to optimize a smaller set of parameters, which improves the success probability.

We further evaluate the possible dependence of these findings on the optimization procedure by comparing different optimizers.
\Fref{fig:optimizer_compare} shows the average success probability for the COBYLA, BFGS, and SPSA optimizers for formulas with clause-to-variable ratio $\alpha = 4.5$.
\Fref{fig:optimizer_compare:bfgs} and \Fref{fig:optimizer_compare:spsa} show that the findings also hold up for different optimizers: For $p>2$, the success probability for \variant{1-3} is higher than that of the baseline implementation.
Furthermore, \Fref{fig:optimizer_compare} shows that SPSA obtains a higher success probability than both COBYLA and BFGS for this experiment.

\begin{figure}[t]
    \begin{subfigure}{1\textwidth}
        \refstepcounter{subfigure}\label{fig:retry_compare:cobyla}
        \refstepcounter{subfigure}\label{fig:retry_compare:bfgs}
        \refstepcounter{subfigure}\label{fig:retry_compare:spsa}
    \end{subfigure}%
    \centering
    \includegraphics{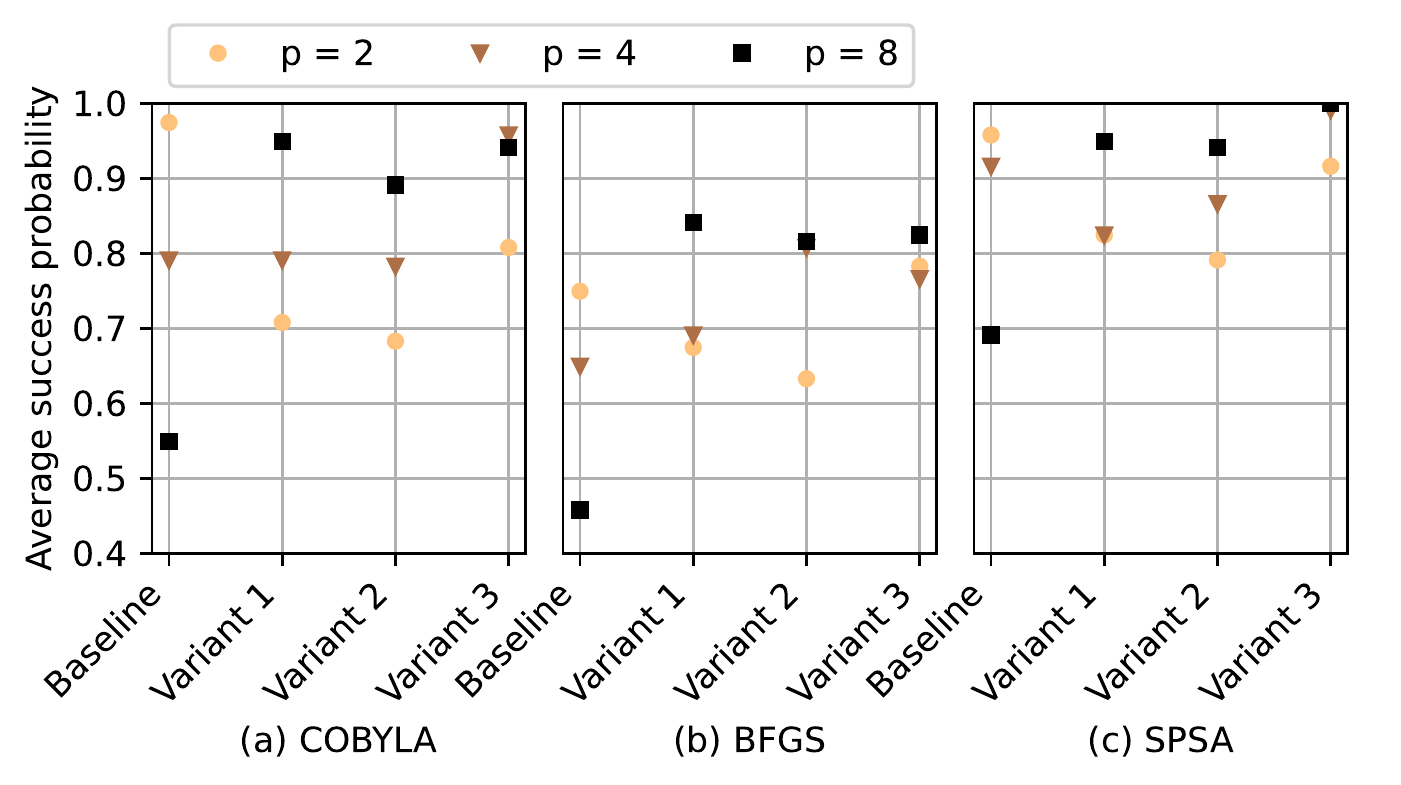}
    \caption{The success probability for the examined variants \threesat formulas with $\alpha = 4.5$ when using five random initial points for optimization and selecting the best result.}
    \label{fig:retry_compare}
\end{figure}

Lastly, \Fref{fig:retry_compare} compares the success probability of the different variants and of the baseline implementation when the optimization loop is repeated with different sets of randomly generated initial parameters.
The baseline implementation even outperforms the examined variants for $p=2$ and the COBYLA optimizer (see \Fref{fig:retry_compare:cobyla}). 
However, already for $p=4$, its success probability decreases.
This suggests that the ansatz that is used in the baseline implementation is expressible enough such that \threesat can be solved with high success probability. 
However, the chance of finding parameters that minimize the cost function for the baseline implementation depends on the initial set of parameters.
With five repetitions, as it is done in \Fref{fig:retry_compare}, the chance to choose initial parameters that simplify the optimization step increases.

Although the success probability for $\variant{1-3}$ also increases when multiple optimization rounds are used (\Fref{fig:retry_compare}), the success probability for \variant{1-3} is less affected by the initial set of parameters. It is already greater than $0.75$ for the COBYLA optimizer and $p=8$ even with only a single round of optimization (\Fref{fig:optimizer_compare:cobyla}).

\subsection{Effect on ansatz complexity}\label{sec:res_ans_comp}

Although the examined changes to QAOA improve the success probability for the \threesat problem, 
some of these adaptions come with an increase in circuit complexity.
Compared to the used bit-flip oracles for the \threesat problem in \emph{Variant~1-3}, the phase-separation operator 
required for the baseline \maxthreesat ansatz can be constructed with linear depth 
using rotations with at most three-qubit interactions (see \ref{sec:app1}
for details). 

The most straightforward approach to obtaining a bit-flip oracle 
for the \threesat problem is by using ancilla qubits to 
store the evaluation result of each clause. This result is obtained 
using single qubit negation and controlled negation gates using three control qubits. 
The result for the overall formula is then obtained using another CNOT gate that is 
controlled by all $m$ ancilla qubits. 
Nielsen and Chuang~\cite{nielsen2002quantum} show how such a gate can generally be 
constructed using a linear number of Toffoli gates and ancilla qubits. There further 
are implementations of this controlled negation that, although they use more operations in total, require no ancilla qubits and 
have the same asymptotic complexity~\cite{Gidney2015}.
Furthermore, there are more sophisticated approaches to constructing the bit-flip oracle for satisfiability problems~\cite{Schmitt2021, Soeken2019, Yang2021}, yet they still require a 
nonconstant number of ancilla qubits. 
Lastly, also the Grover mixer used in \variant{3} increases the ansatz complexity compared to the 
transverse-field mixer $H_B$. However, \Eref{eq:reform_grover_mixer_layers} shows that 
this increase in complexity is not as costly as for the cost operator:
Apart from a phase gate controlled by $n-1$ qubits with linear depth, only two layers each of $H$ and $X$ 
gates have to be used. 

To summarize, the adapted mixer in \variant{3} uses a linear amount of gates as 
opposed to the constant number of gates in \variant{1} and \variant{2}. The 
asymptotic complexity of the ansatz circuit is dominated by the phase-separation operator with linear depth.
Therefore, the asymptotic complexity of the overall ansatz circuit does not increase when compared 
to the implementation of QAOA for the \maxthreesat problem. But since the 
examined circuits are small for the studied instances, this still amounts to 
a considerable increase of gate operations.
Therefore, although the results show 
that the success probability can be improved, this improvement comes with the 
tradeoff of increased circuit complexity. However, the experiments suggest that 
the presented variants still are useful adaptions. Especially \variant{1} and \variant{2} 
increase the success probability using only the 
relatively cheap adaption for the phase-separation operator.

\section{Related Work}\label{sec:relwork}

The performance of the baseline implementation of QAOA for 
\maxthreesat, when applied to the \threesat problem, was 
extensively studied by Zhang et al.~\cite{Zhang2021}. Their examination 
shows that the phase transition in the satisfiability probability 
impacts the success probability for the \maxthreesat problem among other related problems.
Their implementation contrasts the experiments performed in this paper since 
they make use of an incremental initialization strategy 
for the variational parameters and select the best result after multiple 
executions of the optimizer.

The modified phase-separation operator in \variant{1} has also 
been employed in a similar way by Jiang et al.~\cite{Jiang_2017}. They 
also use a cost Hamiltonian that solely distinguishes satisfying 
states from unsatisfying states. In contrast to our 
work, they use it to solve a 
problem more closely related to Grover's original formulation, where 
there is only one possible solution state that should be amplified.
Furthermore, they make use of the transverse-field mixer and fix 
its parameter $\beta$ 
to $\pi$. 

To tackle the problem of variational quantum algorithms amplifying 
suboptimal solutions, 
Bennett and Wang \cite{Bennett2021} employ a similar adaption to \variant{1} 
to the more general 
framework of the Quantum Walk Optimisation Algorithm (QWOA). They present 
the Maximum Amplification Optimization Algorithm (MAOA) and apply it to 
combinatorial optimization problems. They use a similar distinction between 
\emph{good} and \emph{bad} solutions to classify the elements in 
the set of feasible solutions into ones meeting a certain threshold for the cost 
function and those that do not. They further show that the variational 
optimization process can be eliminated as the maximal amplification of the 
solutions state is obtained by a known set of parameters.

In \variant{3}, the Grover mixer is introduced, which 
closely follows the diffusion operator used in amplitude amplification. 
This operation is essential to the variational reformulation of the 
Grover search algorithm~\cite{Morales_2018, Akshay2020reachability} and is further used in~\cite{Bartschi_2020}, who study the application 
of QAOA using this mixer on a range of optimization problems.
They, however, focus on using this mixer on \emph{constrained}
optimization problems, where the set of feasible states is only a subset 
of the set of all possible states that are examined in our experiments.

Another approach that is closely related to \variant{3} is Threshold QAOA~(Th-QAOA)~\cite{Golden_2021}.
Herein, the objective function is replaced by a similar operator 
based on a threshold. It is applied to optimization problems 
and distinguishes states with a cost larger than 
the given threshold from states with less or equal cost.
They further combine this phase-separation operator with the Grover mixer~\cite{Bartschi_2020}
(GM-Th-QAOA)
and evaluate the approximation ratio for different optimization tasks. They 
show that their approach consistently outperforms the standard QAOA using 
the Grover mixer.

The approaches listed here are usually presented in isolation and 
proofs of their complexity are presented or properties of their behavior are analyzed. 
Furthermore, many of the algorithms are proposed for optimization problems. 
This contrasts our work in that we aim to build upon these approaches by
experimentally investigating the effects of these adaptions 
when applied to decision problems such as \threesat specifically.

Lastly, the variants of QAOA we examined in this work 
are closely related to the process of Grover adaptive search~\cite{Durr1996, Baritompa2005}.
This algorithm repeatedly performs the Grover search to only amplify 
states that improve the cost relative to a currently found optimum,
while simultaneously varying the number of iterations of the quantum circuit.
As such it is an adaption of Grover's algorithm to solve 
optimization problems.

\section{Discussion}\label{sec:disc}

In contrast to existing approaches, where QAOA has only been used to solve the 
\maxthreesat optimization problem and the approximated solution is used to 
infer the decision for \threesat, we adapted QAOA to solve the decision problem 
itself to improve the success probability.

The results summarized in \Fref{fig:avg_probs_overall} show, that 
the success probability increases for all examined variants of QAOA 
compared to the baseline implementation. In addition to the 
slight improvement obtained by the adapted cost function 
in \variant{1} for larger $p$, presetting half of the required 
parameters as it is done in \variant{2} is a worthwhile adaption. 
\variant{2} both improves the success probability compared to \variant{1}
and reduces the complexity of the variational optimization due to the 
decreased number of required parameters.
Further investigations are needed to evaluate whether using these 
parameters not as a fixed point but as initial values for 
a warm-started~\cite{Egger_2021, Truger_2022} version of this solver is beneficial.

The adaption that improves the success probability the most 
is \variant{3}, which uses the Grover mixer. This suggests that the 
reason why the standard version of QAOA amplifies
suboptimal solutions is that these solutions are reached more easily with the 
transverse-field mixer $H_B$. Using the adapted mixer $H_G$, this source of errors 
is somewhat mitigated as the experiments show. From the 
experiments conducted in this work, it is, however, not clear whether this improvement 
can be generalized to other decision problems, as the difference in their cost functions might influence this phenomenon.

Overall, one component of the algorithm that limits 
the performance is the classical optimization routine. 
This is especially notable for the results of the baseline implementation 
in \Fref{fig:avg_probs_overall}. The success probability for 
the \maxthreesat-based solver decreases with increasing 
QAOA depth. This is consistent with observations made by Truger~et~al.~\cite{Truger_2022}
that the increased dimension of the parameter space negatively influences 
the optimizer performance. 
To improve the quality of solutions obtained 
by the baseline implementation, techniques such as incremental optimization 
of the parameters~\cite{Truger_2022} can be used or different classical
optimizers can be applied.

Our comparison in \Sref{sec:compo_opt} showed that although the success probability varies when other optimizers are used, \variant{3} still obtains the highest success probability for all optimizers. 
The examination of the success probability when multiple runs per formula are performed also gives an indication of why the presented variants outperform the \maxthreesat-based solver: \emph{Variants 1-3} are less sensitive to the randomly chosen initial set of parameters for the optimization. 
Restarting the optimization procedure multiple times for the \maxthreesat-based solver, increases the chances of obtaining starting parameters that minimize the cost function.
This improves the success probability for $n=10$.
However, as instance sizes grow, the number of required variational parameters grows~\cite{Akshay2020reachability} and the chance to obtain such favorable starting parameters decreases. 
For this reason, the lower sensitivity to the quality of the initial parameters in $\variant{1-3}$ is advantageous.

For practical usage of the algorithm, measuring only one 
satisfying assignment, as it is done in the presented experiments,
suffices for proving that a given formula is satisfiable. 
Given the small instance size of $n=10$, 
this could result in solutions that are found only by chance due to the relatively large number of 
circuit measurements conducted in the variational optimization. 
\begin{figure}[t]
    \centering
    \includegraphics[width=1.0\textwidth]{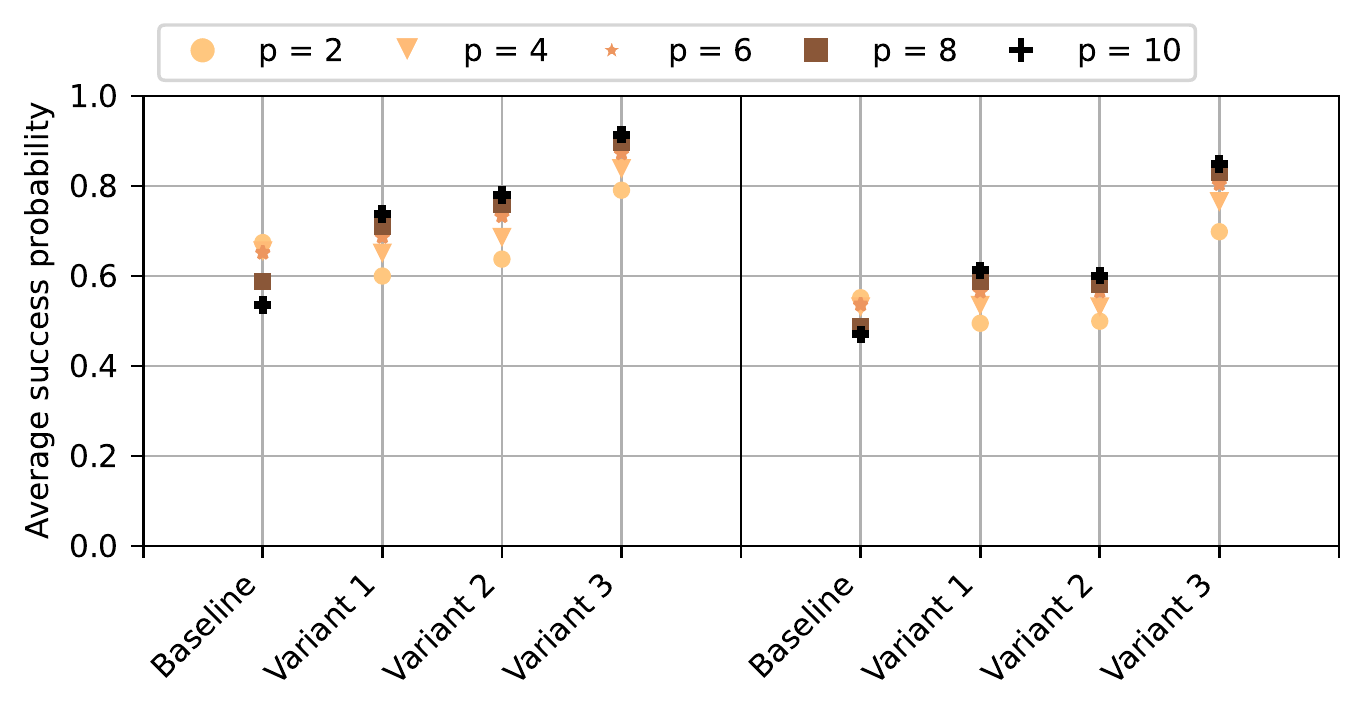}
    \hspace*{\fill}
    \begin{subfigure}[b]{0.5\textwidth}\caption{$T=0.10$}\label{}\end{subfigure}
    \begin{subfigure}[b]{0.4\textwidth}\caption{$T=0.25$}\label{}\end{subfigure}
    \caption{The success probability for the examined variants and the baseline implementation for 
    a satisfiability threshold of (a) $T=0.10$ and (b) $T=0.25$.}
    \label{fig:exp_threshold}
\end{figure}
To address the question of whether results from our experiments 
also hold for larger problem instances, we additionally examine the rate of 
amplification of satisfying assignments. To accomplish this,
a threshold $0 < T \leq 1$ for the proportion 
of measured satisfying assignments before a formula is deemed satisfiable is introduced. 
We reevaluate the experiment results by assuming that a formula is satisfiable 
if and only if
\begin{equation}
    \braket{\vec{\gamma}^*, \vec{\beta}^*|\hthreesat|\vec{\gamma}^*, \vec{\beta}^*} < (1 - T).
\end{equation}
\Fref{fig:exp_threshold} shows the comparison of the different variants using this 
threshold criteria for $T=0.10$ and $T=0.25$. The improvement from \variant{1} to 
\variant{2} becomes less pronounced as the threshold grows. 
However, it shows that the adaptations presented in this paper 
still improve the success probability for the \threesat problem.
Particularly, \variant{3} still shows the largest increase in success probability, even with increasing~$T$. 
This suggests that the improvements were not observed solely because 
of the large number of circuit executions. This evaluation supports the 
finding that the presented adaptions of QAOA improve the algorithm even 
for larger problem instances.

The results in \Fref{fig:exp_standard} exhibit a similar dependency on the clause-to-variable ratio 
as has already been shown for other classical algorithms~\cite{Cheeseman1991, Gent94thesat, Kilby2005}
as well as other quantum algorithms for the \threesat problem~\cite{Akshay2020reachability, Zhang2021}.
In the classical context, usually, the computational resources such as the number of steps
required to solve \threesat instances is studied. However, 
the decrease in success probability of the studied quantum algorithms exhibited in this work 
can be regarded as the same phenomenon: Since 
a reduced success probability directly implies that a deeper QAOA ansatz needs to be 
used, it again results in an increase in needed resources.
For a larger number of variables, the exact point of lowest success probability might deviate from 
the presented results. 
As \Fref{fig:phase_transition} shows, the change in satisfiability probability in the 
phase transition region 
for randomly generated inputs becomes more and more pronounced with an increased 
number of variables. This results in the point of $50\%$ satisfiability probability
being reached at a lower clause-to-variable ratio.
Therefore, we suspect, that the position of minimal success probability 
might also be reached at a lower clause-to-variable ratio for increased instance sizes
for the studied variants of QAOA.

To summarize, the presented experiments show that the success probability of QAOA 
for random \threesat instances can be improved by employing adaptions from 
amplitude amplification algorithms as it is done in \variant{1-3}. These 
improvements come with the cost of increased circuit depth. However, as \variant{1} and 
\variant{2} show, the success probability can be improved even with a moderate increase 
in circuit depth.
Furthermore, the results are valid even when a more strict threshold is used, 
highlighting the usefulness of these adaptions for QAOA.

\section{Summary and future work}\label{sec:conc}

Solving the \threesat problem using QAOA is hindered by the algorithm's 
tendency to amplify suboptimal solutions which decreases its success 
probability. In this work, we introduced three variants of
QAOA that make use of concepts from 
amplitude amplification to alleviate this problem. The variants were 
experimentally examined for their potential to improve the success probability 
of QAOA for the \threesat problem. 
We showed that 
a binary cost function for the phase-separation operator as well as 
the application of the Grover mixer
improves the results for randomly generated \threesat instances.
This improvement was also observed when different optimizers were compared.
Furthermore, using these approaches, the number of parameters to be optimized 
is halved which reduces the complexity for the classical optimizers 
which in turn further improves the success probability.

To investigate whether these findings also hold up for 
larger problem instances, we examined the rate of amplification of 
satisfying variable assignments by using a threshold for satisfiability
in the samples obtained by the ansatz. The improvement in the success 
probability is still noticeable. However, the change in the parameter 
landscape that is introduced by the reformulation of the cost function 
might still impact the results for larger instances. This effect 
remains to be fully investigated in future work.

The adaptions examined in this work increase the 
depth of the ansatz circuit. This tradeoff has to be carefully considered when 
choosing which adaptions of the QAOA ansatz to use for specific problems.
Furthermore, the effect the chosen classical optimization procedure has on 
the result also needs to be taken into account and it 
remains to be studied in future work, whether 
these results hold even in the presence of noise. This can either be accomplished by performing 
the experiments on real quantum devices, given that instance sizes can be made 
sufficiently small to allow such experiments, or by introducing a noise model 
into the simulation procedures.

In future work, we propose to examine if other common optimizations of QAOA
such as warm-starting can further improve the success probability 
for \threesat.
Furthermore, various other decision problems, such as 
finding monochromatic triangles or finding exact set covers
might benefit from the 
adaptions presented in this work and should be considered in future work as well.

\ack 
This work was partially funded by the BMWK projects \textit{PlanQK} (01MK20005N) and \textit{EniQmA} (01MQ22007B).

\appendix
\section{\maxthreesat phase-separation circuit}\label{sec:app1}

The cost Hamiltonian for \maxthreesat (see also the supplementary material of~\cite{Zhang2021}) is given as 
\begin{equation}
    \hmaxthreesat = \sum_{j=1}^m \frac{1}{8} \prod_{i=1}^3 (I_{a_{i,j}} + p_{i,j} Z_{a_{i,j}}).
\end{equation}
In the following equations, we omit the polarity of the literals and assume $p_{i,j} = 1$, 
since this does not influence the upper bound on the number of gate operations that have to be performed per clause.
The product for one clause can be expanded 
as 
\begin{eqnarray}
    \left(I_1 + Z_1\right)\left(I_2 + Z_2\right)\left(I_3 + Z_3\right)\\
    = I + Z_1 + Z_2 + Z_3 + Z_1Z_2 + Z_1Z_3 + Z_2Z_3 + Z_1Z_2Z_3
\end{eqnarray}
where we assume for simplicity, that the atomic variables $a_1$, $a_2$ and $a_3$ participate in the clause.
Therefore, one clause can be implemented as 
\begin{equation}
    e^{-i \gamma H_{C_j}} = \prod_{k=1}^3 \left[ e^{-i \gamma Z_k} \cdot \prod_{0<l<k} e^{-i \gamma Z_l Z_k} \right] \cdot e^{-i \gamma Z_1 Z_2 Z_3},
\end{equation}
which results in three RZ rotations, three RZZ rotations and one three-qubit interaction 
resulting in an RZZZ gate~\cite{Zhang2021}. Overall, this shows
that the subcircuit for one clause is of constant depth, resulting in a 
linear-depth phase-separation operator.

\section*{References}
\bibliography{mm}

\end{document}